\documentstyle[aps,pre,floats,epsf,draft]{revtex}

\begin{document}

\twocolumn[{\hsize\textwidth\columnwidth\hsize\csname
@twocolumnfalse\endcsname
\title{
\draft Comment on ``Mechanisms of synchronization and pattern
formation in a lattice of
pulse-coupled oscillators''
}
\author{
Jos\'e M. G. Vilar
}
\address{Departments of Physics and Molecular Biology,
Princeton University, Princeton, NJ 08544
}
\maketitle
\widetext
\begin{abstract}
In a recent paper, Diaz-Guilera {\it et al.}
[Phys. Rev. E {\bf 57}, 3820 (1998)]
analyze the mechanisms of synchronization and pattern formation in a 
lattice of pulse-coupled oscillators.
In essence, their analysis consists in the study of the stability of the fixed
points of several linear return maps which are obtained from the original
system by means of matrix manipulations.
We show that although the model they consider
is very specific and actually unable to account even
for  a linear phase response curve, their
method does not give correct information on the original system
since many of the assumptions involved
are in general not correct. To clarify these aspects,
several issues concerning the real dynamics are also discussed.
\end{abstract}

}]

\narrowtext

\section{Introduction}

In Ref.~\cite{ladi}, Diaz-Guilera {\it et al.} propose a new analytical
procedure to study a lattice model of pulse-coupled
oscillators in a one-dimensional ring with unidirectional coupling.
This procedure, which consists in  the linear stability analysis of
the fixed points of some return maps obtained from the original
system by means of matrix manipulations, is based on some
{\it ad hoc} assumptions and in a particular form for the phase
response curve (PRC).
Diaz-Guilera {\it et al.} claim that their description gives
complete information on the original system as well as that the results
they obtain for the particular PRC they consider can also be
extrapolated to a general PRC.
In this comment we show that the model they consider
is very specific and unable to account even for the case of a linear PRC.
We also show that Diaz-Guilera {\it et al.}'s results, in addition,
do not give complete information on the original system, not even
for the specific PRC they
consider, owing to the fact that many of the assumptions involved
are not valid.

A first step toward simplifying the system and getting a linear description
is to consider a linear PRC. This requirement
for the PRC is often not too restrictive since one can expand the
PRC in powers of the convexity of the driving hopping to grasp the behavior
of the system by only considering the first two terms in the expansion,
i.e. a constant term plus another proportional to the phase.
Diaz-Guilera {\it et al.}, in contrast, only consider the term proportional
to the phase and claim that the constant term is unimportant
since its only effect is to shift the threshold. This claim, which
is not proved neither in Ref.~\cite{ladi} nor in the reference they quote,
is not true. This fact can easily be seen, for instance,
in the inhibitory situation if one considers the linear PRC,
$\Delta(\phi)=\alpha+\epsilon\phi$, with  $-1-\alpha \le \epsilon < -\alpha$
and $\alpha < 0$.
In this case, when an oscillator
gets a pulse it can or cannot instantaneously be reset to zero depending
on the value of its phase, $\phi$.
In the case addressed by Diaz-Guilera {\it et al.}, when the interactions
are  inhibitory, an oscillator can never be reset to zero after
getting a pulse if $\epsilon>-1$ or it is always reset
to zero if $\epsilon=-1$, i.e., whether an oscillator is 
reset to zero or not does not depend on the value of its phase.
This simple example makes it evident that the approximation of Diaz-Guilera
{\it et al.} is not only unable to account even for the general linear case
but also that it is highly pathological.
In this sense, one should hardly expect that the
results they obtain could be extrapolated to any other system with some
degree of generality, as we will show.

\section{Negative coupling}

To proceed further with their analysis, Diaz-Guilera {\it et al.}  assume
also that the oscillators fire in a cyclic order, i.e.,  advancements
between oscillators are not allowed. This additional assumption,
which holds in the all-to-all case, enables
them to construct some linear return maps for the particular PRC they
are considering. 
These return maps are intended to describe the 
dynamics of the system 
and since they are linear, it can be done easily
by only looking at their fixed points and the corresponding eigenvalues.
In their analysis, however,
Diaz-Guilera {\it et al.} additionally constrain the fixed points
to those in which each oscillator fires just once,
neglecting then other possibilities.
For the case in which $\epsilon$ is negative they found that
the fixed points are stable whereas they are unstable for the case
in which $\epsilon$ is positive. 
After performing some simulations
for a system of three oscillators they realize that in fact advancements
are possible but they claim, again without proof, that advancements only
matter during the transient dynamics. Thereby, they proceed with their
analysis without taking into account this annoying hindrance.  

\begin{figure}[t]
\centerline{
\epsfxsize=8.0cm 
\epsffile{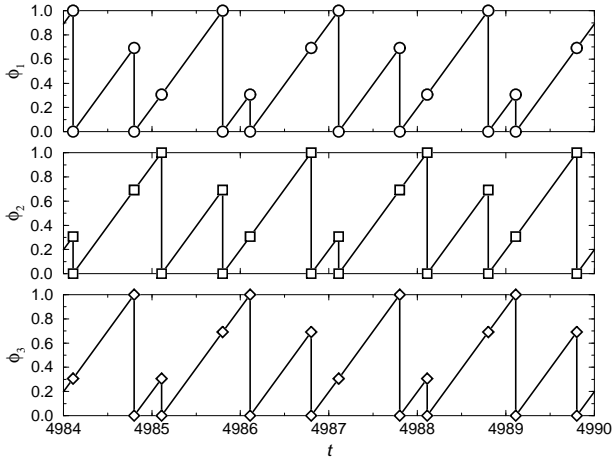}
}
\caption[a]{\label{fig1}
Time evolution for a lattice of three oscillators
for Diaz-Guilera {\it et al.}'s
PRC with $\epsilon=-1$.
The symbols are depicted each time any oscillator reaches the
threshold.
}
\end{figure}

In Fig. \ref{fig1} we have displayed the typical time evolution for the 
Diaz-Guilera {\it et al.}'s PRC for the case of three oscillators
when the phase of an oscillator is
always reset to zero after receiving a pulse.
For this coupling,  Diaz-Guilera {\it et al.}'s
results predict that there exists a stable fixed point in which just before
an oscillator fires, the other two should have a phase equal to $0.5$.
Fig. \ref{fig1}, however, clearly illustrates that such a state is not reached.
Diaz-Guilera {\it et al.} incorrectly computed the moduli
of the eigenvalues of the $2\times2$ matrix for $\epsilon<-3/4$.
The eigenvalues
are $0$ and $-1$ whose moduli are not $0$ [$=(1+\epsilon)^{3/2}$],
as they obtained. In this case the fixed point of the return map they
construct is not stable despite $\epsilon$ being negative.
Actually,  for arbitrary initial conditions,
$\{\phi_1^*,\, \phi_2^*,\, \phi_3^*\}$,
where as a first oscillator has been taken
the one with the highest phase,
the system  eventually attains a state in
which just before the first oscillator reaches the threshold, the phases
of the other two oscillators alternate between $\phi_2=\phi_1^*-\phi_3^*$,
$\phi_3=\phi_1^*-\phi_3^*$ and $\phi_2=1-(\phi_1^*-\phi_3^*)$,
$\phi_3=1-(\phi_1^*-\phi_3^*)$. Notice that although
Diaz-Guilera {\it et al.}'s solution
is a fixed point of the real dynamics, the set of initial conditions
for which the system reaches the corresponding state has zero measure
since $\phi_1^*-\phi_3^*$ must be equal to $0.5$. In essence, this fact
means that given arbitrary
initial conditions the probability that Diaz-Guilera {\it et al.}'s results
describe the eventual evolution of the system is zero.
In this case,  the real attractor of the dynamics is not 
precisely a fixed point of the ``one-cycle'' return map, 
but a state of period  two, which in addition depends 
on the initial conditions.

\begin{figure}[t]
\centerline{
\epsfxsize=8.0cm 
\epsffile{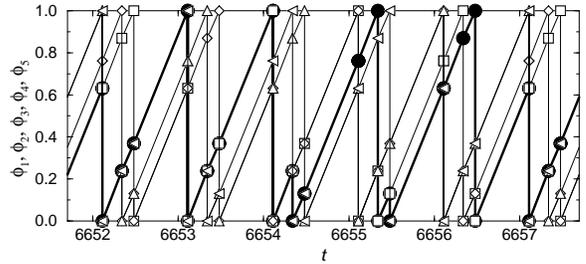}
}
\caption[a]{\label{fig2}
Time evolution for a lattice of five oscillators
for Diaz-Guilera {\it et al.}'s PRC with $\epsilon=-1$.
Filled circles, squares, diamonds, triangles up, and triangles
left correspond to $\phi_1$, $\phi_2$, $\phi_3$, $\phi_4$,
and $\phi_5$, respectively. The symbols are depicted each time
that any oscillator reaches the threshold.
}
\end{figure}

The failure of Diaz-Guilera et. al.'s scheme to account even
for the specific model they are considering when phase
resettings are allowed,
seems to be a general feature for any number of oscillators.
An example of this behavior is illustrated
in Fig. \ref{fig2}, where we consider the case of five oscillators.
This time series clearly does not agree with 
Diaz-Guilera {\it et al.}'s predictions.
At first glance, one could wonder if  that time series is just
a long transient. It is easy to see, however, that a periodic
state in which each oscillator fires four times,
not just once as Diaz-Guilera {\it et al.}'s procedure imposes, is reached.
Notice that the phases after $5$ time units,
for instance at $6652$ and $6657$, have the same value.
In addition, the figure shows that the order in which the oscillators
reach the threshold is not preserved.
This result indicates that the assumption of the oscillators
firing in a cyclic order except perhaps in the transient,
as assumed by  Diaz-Guilera {\it et al.}, is in general not valid
and other kind of behavior with different periodicity appears.
From the methodological point of view, the previous examples
illustrate that the existence of
advancements cannot be disregarded.

It is worth to notice that although in the appendix
Diaz-Guilera et. al. conclude
that for an arbitrary number of oscillators
the moduli of the eigenvalues are always
lower than $1$ when $\epsilon <0$, in fact,
if the moduli are correctly computed,
that result is in general not valid.
For instance, in the case of three oscillators there is
an eigenvalue with modulus $1$ when $\epsilon=-1$.
We performed numerical simulations for the same PRC as in
previous figures and for values
of the number of oscillators ranging from $3$ to $1000$ and in all of them
we found that Diaz-Guilera et. al.'s results were unable to account
for the real dynamics.
In the opposite situation, when we considered values for the coupling
which never reset the phase to zero, we obtained results
that can be compatible with Diaz-Guilera et. al.'s predictions.

\section{Positive coupling}

In the case wherein the coupling is excitatory
($\epsilon > 0$),
by only taking into account the results
concerning the stability of
the previous fixed points and by assuming, again without proof,
that a set of synchronized oscillators acts as a single unit that
cannot be broken, Diaz-Guilera {\it et al.} concluded that 
eventually the whole population fires in  synchrony.
This procedure to state the presence of synchrony
is clearly inconsistent with the assumptions they use
to try to describe the real system through linear return maps.
They assumed that advancements are only important during the transient
dynamics. However, the whole time evolution until
synchrony is reached is precisely a transient. In addition,
the fact that the state corresponding to the cyclic ordering
they propose for the eventual evolution is unstable does not imply
that the system will
go far away from those fixed points.
For instance, the system might oscillate
around an unstable fixed point as occurs in the
logistic map, as well as in may other systems, when period doubling
appears~\cite{doubling}. 

Notice that Figs. \ref{fig1} and \ref{fig2} show also that,
in contrast to the all-to-all case, when short-range
interactions are present two synchronized oscillators
can lose their mutual synchrony.
This situation is  a general
property of the model, even for the excitatory case~\cite{VC}.
In general, 
there are not ``absorbing barriers surrounding the repellers'',
as quoted by Diaz-Guilera {\it et al.}
In the real dynamics the situation is more complex: 
Two synchronized oscillators can lose
their mutual synchrony but eventually they are able to recover it
if some conditions for the PRC are fulfilled~\cite{VC}.

It is fair to say that sufficient conditions for
synchronization to occur in the all-to-all case were rigorously
found by Mirollo and Strogatz~\cite{Mirollo}. Based on numerical
simulations,
these authors also conjectured that in the excitatory case
the same sufficient conditions should hold if a local coupling
is considered. 
In this regard, despite the procedure followed
by Diaz-Guilera {\it et al.} in view to show synchrony is not correct,
their model meets the criteria for synchronization to occur
when $\epsilon$ is positive~\cite{VC}. 
In essence, Diaz-Guilera {\it et al.} results concerning
synchronization are another numerical verification of 
Mirollo and Strogatz's  conjecture~\cite{Mirollo}, which has
been widely analyzed through numerical
simulations~\cite{Corral} and recently rigorously proved in a 
model with short-range interactions~\cite{VC}.

\section{Linear coupling}

\begin{figure}[t]
\centerline{
\epsfxsize=9cm 
\epsffile{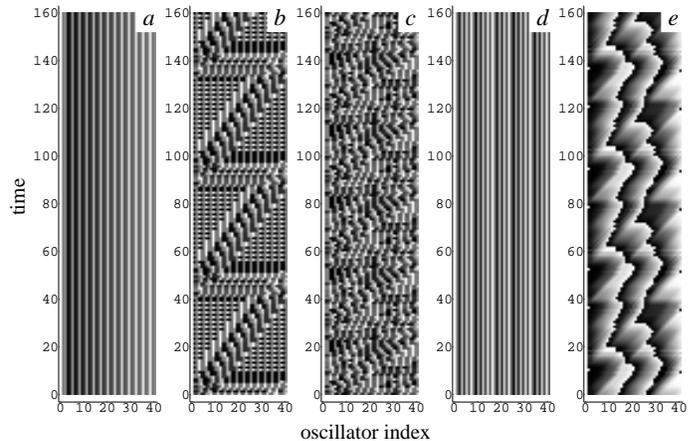}
}
\caption[a]{\label{fig3}
Spatiotemporal evolution corresponding to a lattice of $41$ oscillators
with $\Delta(\phi)=\alpha+\epsilon\phi$ for
(a) $\alpha=-0.1$, and $\epsilon=-0.6$,
(b) $\alpha=-0.2$, and $\epsilon=-0.6$,
(c) $\alpha=-0.1$, and $\epsilon=-0.8$,
(d) $\alpha=0.3$, and $\epsilon=-0.3$, and
(e) $\alpha=-0.3$, and $\epsilon=0.3$.
The horizontal direction represents the oscillator index
and the vertical direction represents time. All the phases have been
displayed just before the first oscillator fires. Black and white colors
stand for high and low values of the phase, respectively.
}
\end{figure}

Previously, we explained how the PRC that Diaz-Guilera {\it et al.}
considered is unable to account even for  a linear PRC
and why one should hardly expect that the
results they obtain can be extrapolated to any  system with a certain
degree of generality. We also showed that the results they obtain
do not account neither for the specific PRC they consider when
the phase is reset to zero after receiving a pulse. 
In order to address the issue of what happens when such pathologies
are not present, we now consider a PRC including the constant term 
neglected by Diaz-Guilera {\it et al.}
In this case, whether the phase is reset to zero or not
will depend on the value of the phase.
When the phase is reset to zero, i.e.,
$\alpha+\epsilon\phi \le -\phi$,
the PRC is effectively described by the $\epsilon=-1$ case 
previously studied since the resulting phase cannot be lower than zero
by definition of the model. In contrast, when the phase is not
reset to zero it is given by $\alpha+\epsilon\phi$.
The existence of these two effective contributions
on the PRC breaks the linearity of the coupling and
makes Diaz-Guilera {\it et al.}'s description inapplicable in a general case.
Three examples of what happens under these circumstances are displayed
in Figs. \ref{fig3}(a), \ref{fig3}(b) and  \ref{fig3}(c). We show the 
time evolution of the return map corresponding
to a decreasing inhibitory PRC ($\alpha<0$ and $\epsilon<0$) for 
three different values of this couple of parameters.
Although Diaz-Guilera {\it et al.}'s procedure does not account even for 
this PRC, Fig. \ref{fig3}(a) 
looks like their predictions,
i.e, the phase of each oscillator is always the
same just before the first oscillator fires. 
However, Figs. \ref{fig3}(b) 
and \ref{fig3}(c) 
are clearly incompatible with their results
since there is not an eventual cyclic firing
and continuous overtakings between oscillators occur. Notice
that these states are not transient since they are periodic.

In Figs.\ref{fig3}(d) and \ref{fig3} (e) we display the
time evolution for an excitatory decreasing and an inhibitory
increasing PRC, respectively.
Although the PRC corresponding to Fig. \ref{fig3}(d) is positive
and in Fig. \ref{fig3} (e) $\epsilon>0$, the system does not synchronize.
These results again do not fit Diaz-Guilera {\it et al.}'s
predictions wherein for a positive PRC
($\epsilon>0$) the system will eventually synchronize.
In general, the appearance of synchronization
cannot be stated by only considering
the inhibitory or excitatory character of the coupling neither
by only considering the derivative of the PRC. In the all-to-all case,
whether the coupling is excitatory or inhibitory,
synchronization appears for a positive derivative of the PRC.
In general, the ``absorbing barriers surrounding the
repellers'' called for by Diaz-Guilera {\it et al.}'s to assert
the existence of synchronization do not exist and the system
will not synchronize despite the interactions be excitatory or the derivate
of the PRC ($\epsilon$) be
positive. When short-range interactions are considered both
conditions are simultaneously required~\cite{VC}.
Synchronization in Diaz-Guilera {\it et al.} simulations appears since
the two conditions are same due to the specific PRC they consider.

\section{Conclusions}

To summarize, we have shown that a general, complete, and correct
description of a lattice model of pulse-coupled oscillators
is not possible trough the method
proposed by Diaz-Guilera {\it et al.}
In essence, Diaz-Guilera {\it et al.}'s results are only valid
when studying the linear stability of the fixed points corresponding
to a cycle in which each oscillator fires exactly once per period,
for a specific PRC which is proportional to the phase and for the
inhibitory situation when phase resettings to zero are not allowed.
The failure for their description to be applied to any other situation
relies on the fact that many of the assumptions required to
the development of their analysis are not valid.
In particular, a PRC proportional to the phase is
not equivalent to a linear PRC; advancements between oscillators 
can be important for the final state and not only during the
transient dynamics; fixed points in which an oscillator fires several
times per period are present; synchronized oscillators can
lose their mutual synchrony. Diaz-Guilera {\it et al.} method can never
be used to analyze the global dynamics of the system since it consists
only in the study of the linear stability of some fixed points.
Therefore, the appearance of synchronization,
which is the most relevant case from the
experimental point of view~\cite{Bas}, can not be inferred from
their results.

Our analysis makes it clear that a priori
indiscriminate assumptions in view to obtain a known result
are not only unjustified but can  also
give a misunderstanding of what is really happening.
Non-linear systems do not usually behave
in the way one can suspect from the intuition arising
only from some simulations for a particular model.
Conversely, rigorous mathematical results~\cite{VC,Mirollo}
are very useful to understand when
a given behavior should be expected and under which conditions
this behavior can or cannot be extrapolated to other systems.

\section*{acknowledgments}

The illuminating discussions with A. Corral and J. M. Rub\'{\i}
are gratefully acknowledged.

%\onecolumn
%\newpage
%\widetext

\end{document}